\documentclass[aps,twocolumn,notitlepage,preprintnumbers,superscriptaddress,amsfont,graphicx,10pt]{revtex4}
\usepackage{epsfig,amsmath,amssymb}
\usepackage{color}
\usepackage{float}
\usepackage{subfigure}
\usepackage[utf8]{inputenc}

\definecolor{BrickRed}{rgb}{0.9,0.1,0}
\input{tcilatex}
\begin{document}

\preprint{DO-TH 16/03}
\title{The $750\,\text{GeV}$ diphoton resonance in the light of a 2HDM with $%
S_3$ flavour symmetry}
\author{A. E. C\'arcamo Hern\'andez}
\email[Electronic address:]{antonio.carcamo@usm.cl}
\affiliation{Universidad T\'{e}cnica Federico Santa Mar\'{\i}a and Centro Cient\'{\i}%
fico-Tecnol\'{o}gico de Valpara\'{\i}so\\
Casilla 110-V, Valpara\'{\i}so, Chile}
\author{I. de Medeiros Varzielas}
\email{ivo.de@soton.ac.uk}
\affiliation{{\small School of Physics and Astronomy, University of Southampton,}\\
Southampton, SO17 1BJ, U.K.}
\author{E. Schumacher}
\email{erik.schumacher@tu-dortmund.de}
\affiliation{{\small Fakult\"at f\"ur Physik, Technische Universit\"at Dortmund}\\
D-44221 Dortmund, Germany}
\date{\today }

\begin{abstract}
Very recently we proposed a predictive 2 Higgs Doublet Model with $S_{3}$
flavour symmetry that successfully accounts for fermion masses and mixings.
In this letter, motivated by the $750$ GeV Higgs diphoton resonance recently
reported by the ATLAS and CMS collaborations, we modify this model by adding
exotic top partners with electric charge $\frac{5}{3}$ and a electrically charged scalar singlet. These exotic top partners decay into a charged scalar singlet and the SM up type quarks, whereas the charged scalar singlet will mainly decay into SM up and down type quarks. This simple modification enables our model to successfully account for the Higgs diphoton excess at $750\,\text{GeV}$ provided that the exotic quark masses are in the range $[1,2]$ TeV, for $O(1)$ exotic quark Yukawa couplings.
\end{abstract}

\maketitle

\textbf{Introduction}.

Recently, the ATLAS and CMS collaborations reported an excess of events
above the expected background in the diphoton final state \cite%
{ATLAS:2015abc,CMS:2015dxe}. This is very promising as both collaborations
have the excess at an invariant mass of about $750\,\text{GeV}$, and with
local statistical significance of $3.9\,\sigma $ (ATLAS) and $2.6\,\sigma $
(CMS). The confirmation of this excess would constitute proof of physics
beyond the Standard Model (SM). Although the excess may turn out to be a
statistical fluctuation, it is very enticing to consider models that include
in their field content something that can account for a resonance at $750\,%
\text{GeV}$ and thus accounts for the excess of events.

This diphoton final state excess has generated much activity in the
community. Many works have considered the excess of events in a model
independent way, see~\cite{Buttazzo:2015txu,
Franceschini:2015kwy,Ellis:2015oso,
Gupta:2015zzs,Chakrabortty:2015hff,Agrawal:2015dbf,Csaki:2015vek,Falkowski:2015swt,Aloni:2015mxa,Huang:2015evq,Cho:2015nxy, Kanemura:2015vcb}%
. Another study considered the implications of the Higgs diphoton resonance
for the stability of the Higgs, naturalness and inflation \cite%
{Salvio:2015jgu} and \cite{Marzola:2015xbh} considers whether the resonance
could be a Coleman-Weinberg inflaton.

Previous works have studied the resonant production at the LHC of (pseudo-)
scalars coupled to two photons and gluons in the mass region from $30$ GeV
to $2$ TeV \cite{Jaeckel:2012yz}, and of the observable effects of new
scalar particles \cite{deBlas:2014mba}.

A myriad of explanations have been considered since the announcement of the
excess. Adding scalar singlets is a fairly straightforward option, see \cite%
{Fichet:2015vvy,Cao:2015pto,Dutta:2015wqh,Alves:2015jgx,Han:2015dlp,Chang:2015sdy,Antipin:2015kgh,Zhang:2015uuo,Cheung:2015cug,Hernandez:2015hrt}%
, and \cite{Altmannshofer:2015xfo} (which explores several scenarios
including a scalar singlet). The scalar singlet model can be further
extended with extra dimensions \cite{Cai:2015hzc}, or, alternatively, with
additional vector leptoquarks, which also allow to simultaneously explain
the B decay anomalies \cite{Bauer:2015boy, Murphy:2015kag}.

Supersymmetric explanations are possible and the excess has been interpreted
in the context of the Minimimal Supersymmetric Standard Model (MSSM) \cite%
{Chakraborty:2015gyj}, its R-symmetry violating version \cite%
{Allanach:2015ixl} and other supersymmetric extensions \cite%
{Wang:2015kuj,Wang:2015omi}.

Extending the SM gauge symmetry can also explain the excess, as described by 
\cite%
{Cao:2015scs,Boucenna:2015pav,Hernandez:2015ywg,Patel:2015ulo,Cao:2015xjz,Cao:2015apa,Pelaggi:2015knk,Chao:2015nac,Chao:2015nsm,Jiang:2015oms,Dong:2015dxw,Dasgupta:2015pbr}%
, and models based on strongly coupled theories were considered by~\cite%
{Franceschini:2015kwy,Buttazzo:2015txu,Harigaya:2015ezk,Molinaro:2015cwg,Bian:2015kjt,Curtin:2015jcv,No:2015bsn,Matsuzaki:2015che,Kim:2015ron,Son:2015vfl}%
. An extended broken symmetry with an extra Higgs boson and massive vector
bosons can account for the diboson anomaly and the anomalous $t\bar{t}$
forward-backward asymmetry \cite{deBlas:2015hlv}

Models with extra fermions \cite{Basso:2012nh}, in particular vector-like
fermions had been considered before~\cite%
{Bonne:2012im,Moreau:2012da,Aguilar-Saavedra:2013qpa,Angelescu:2015kga} and
after the announcement~\cite%
{Franceschini:2015kwy,Buttazzo:2015txu,Angelescu:2015uiz,Ellis:2015oso,
Kobakhidze:2015ldh,
Falkowski:2015swt,Benbrik:2015fyz,Dhuria:2015ufo,Das:2015enc,Liu:2015yec,Tang:2015eko,Li:2015jwd,Dev:2015vjd}%
, and other works relate the diphoton excess with loop TeV-scale seesaw
mechanisms, either at two \cite{Chao:2015nac} or three loops \cite%
{Kanemura:2015bli}.

Relating the excess with dark matter has been extensively considered, see~%
\cite%
{Mambrini:2015wyu,Backovic:2015fnp,Nakai:2015ptz,Knapen:2015dap,Bian:2015kjt,Martinez:2015kmn, Bi:2015uqd, Bauer:2015boy, Barducci:2015gtd, Dey:2015bur, Dev:2015isx, Han:2015yjk,Park:2015ysf}%
.

There are also sgoldstino~\cite%
{Bellazzini:2015nxw,Petersson:2015mkr,Demidov:2015zqn}, radion ~\cite%
{Ahmed:2015uqt,Bardhan:2015hcr}, graviton~\cite{Arun:2015ubr} and exotic
heavy axion~\cite{Pilaftsis:2015ycr,Harigaya:2015ezk,Molinaro:2015cwg}
interpretations of the $750\,\text{GeV}$ excess.

String-motivated models were considered in \cite%
{Heckman:2015kqk,Cvetic:2015vit,Anchordoqui:2015jxc}.

Finally, a rather natural and popular framework to explain the excess is
that of 2 Higgs Doublet Models (2HDMs), which have been studied in~\cite%
{DiChiara:2015vdm,Angelescu:2015uiz,Becirevic:2015fmu,Han:2015qqj,Badziak:2015zez, Moretti:2015pbj,Bi:2015lcf,Bizot:2015qqo,Kang:2015roj,Huang:2015rkj}
and will also be considered in this letter, where we make a simple
modification of an existing 2HDM \cite{Hernandez:2015dga}, by adding exotic
top quark partners.


\textbf{The model}. We consider the 2HDM that we recently proposed in \cite%
{Hernandez:2015dga}, with the SM gauge symmetry supplemented by the $%
S_{3}\otimes Z_{3}\otimes Z_{3}^{\prime }\otimes Z_{14}$ discrete group. The
scalar sector has two Higgs doublets (assigned as trivial $S_{3}$ singlets)
plus four SM singlet scalars assigned as one $S_{3}$ trivial singlet ($\chi $%
), one $S_{3}$ non-trivial singlet ($\zeta $) and one $S_{3}$ doublet ($\xi $%
). In order to successfully explain the LHC diphoton excess at $750\,\text{%
GeV}$, we extend the fermion sector of our 2HDM by including four $SU(2)_{L}$
singlet exotic quark fields with electric charge $\frac{5}{3}$, $T_{1L}$, $%
T_{1R}$, $T_{2L}$, $T_{2R}$, grouped into two $S_{3}$ doublets, i.e., $%
T_{L}=\left( T_{1L},T_{2L}\right) $, $T_{R}=\left( T_{1R},T_{2R}\right) $.
These exotic quark fields are neutral under the $Z_{3}\otimes Z_{3}^{\prime
}$ discrete symmetry but charged under the $Z_{14}$ symmetry as:

\begin{equation}
T_{L}\rightarrow T_{L},\hspace{0.5cm}T_{R}\rightarrow e^{\frac{\pi i}{7}%
}T_{R}.  \label{TZ14}
\end{equation}

In addition, the instability of the exotic quarks $T_{1L}$, $T_{1R}$%
, $T_{2L}$, $T_{2R}$ requires that we extend the scalar sector with a
single electrically charged SM scalar singlet $\rho ^{+}$. We assume that
$\rho ^{+}$ is a trivial $S_{3}$ singlet and is
neutral under the $Z_{3}\otimes Z_{3}^{\prime }\otimes Z_{14}$ discrete
symmetry. The exotic quarks should then decay 
into $\rho ^{+}$ and SM up type quarks, through the Yukawa interactions given in Eq. (\ref{yuk}), 
while $\rho ^{+}$ in turn will mainly decay into SM up and down type
quarks. The remaining particles have the $S_{3}\otimes Z_{3}\otimes
Z_{3}^{\prime }\otimes Z_{14}$ charge assignments
described in \cite{Hernandez:2015dga}. In this model, the $S_{3}$ symmetry
reduces the number of parameters in the Yukawa sector making this 2HDM more
predictive, and the remaining symmetries control the allowed Lagrangian
terms by distinguishing the fields. For example, the two scalar $SU(2)_{L}$
doublets have different $Z_{3}$ charges ($\phi _{1}$ being neutral). The $%
Z_{3}^{\prime }$ and $Z_{14}$ symmetries shape the hierarchical structure of
the fermion mass matrices necessary to get a realistic pattern of fermion
masses and mixing. The assignments of the scalar and fermion particles
particles are given in \cite{Hernandez:2015dga}, giving rise to the
following Yukawa terms for the quark and lepton sectors:

\begin{eqnarray}
\tciLaplace _{Y}^{q} &=&\varepsilon _{33}^{\left( u\right) }\overline{q}_{3L}%
\widetilde{\phi }_{1}u_{3R}+\varepsilon _{23}^{\left( u\right) }\overline{q}%
_{2L}\widetilde{\phi }_{2}u_{3R}\frac{\chi ^{2}}{\Lambda ^{2}}+\varepsilon
_{13}^{\left( u\right) }\overline{q}_{1L}\widetilde{\phi }_{2}u_{3R}\frac{%
\chi ^{3}}{\Lambda ^{3}}\notag \\
&&+\varepsilon _{22}^{\left( u\right) }\overline{q}_{2L}\widetilde{\phi }%
_{1}U_{R}\frac{\xi \chi ^{3}}{\Lambda ^{4}}+\varepsilon _{11}^{\left(
u\right) }\overline{q}_{1L}\widetilde{\phi }_{1}U_{R}\frac{\xi \chi
^{4}\zeta ^{3}}{\Lambda ^{8}}\notag \\
&&+\varepsilon _{33}^{\left( d\right) }\overline{q}_{3L}\phi _{1}d_{3R}\frac{%
\chi ^{3}}{\Lambda ^{3}}+\varepsilon _{22}^{\left( d\right) }\overline{q}%
_{2L}\phi _{2}d_{2R}\frac{\chi ^{5}}{\Lambda ^{5}}\notag \\
&&+\varepsilon _{12}^{\left( d\right) }\overline{q}_{1L}\phi _{2}d_{2R}\frac{%
\chi ^{6}}{\Lambda ^{6}}+\varepsilon _{21}^{\left( d\right) }\overline{q}%
_{2L}\phi _{2}d_{1R}\frac{\chi ^{6}}{\Lambda ^{6}}\notag \\
&&+\varepsilon _{11}^{\left( d\right) }\overline{q}_{1L}\phi _{2}d_{1R}\frac{%
\chi ^{7}}{\Lambda ^{7}}+y_{T}\overline{T}_{L}T_{R}\chi+\varepsilon_{\rho}\overline{q}_{3L}\phi_{2}T_{R}\frac{\chi\rho ^{-}}{\Lambda^2}\notag  \\
&&+\varepsilon _{33}^{\left( \rho \right) }\overline{q}_{3L}\phi _{1}u_{3R}%
\frac{\rho ^{-}}{\Lambda }+\varepsilon _{33}^{\left( \rho \right) }\overline{%
q}_{3L}\widetilde{\phi }_{1}d_{3R}\frac{\rho ^{+}\chi ^{3}}{\Lambda ^{4}}\notag \\
&&+\varepsilon _{23}^{\left( \rho \right) }\overline{q}_{3L}\widetilde{\phi }%
_{2}d_{2R}\frac{\rho ^{+}\chi ^{3}}{\Lambda ^{4}}+\varepsilon _{32}^{\left(
\rho \right) }\overline{q}_{3L}\phi _{2}U_{R}\frac{\xi \rho ^{-}\chi }{%
\Lambda ^{3}}\notag \\
&&+y_{1\rho }\overline{T}_{L}U_{R}\rho ^{+}\frac{\chi }{\Lambda }+y_{2\rho }%
\overline{T}_{L}U_{R}\rho ^{+}\frac{\zeta ^{3}\chi }{\Lambda ^{4}}+h.c \label{yuk}
\end{eqnarray}

\begin{eqnarray}
\tciLaplace _{Y}^{l} &=&\varepsilon _{33}^{\left( l\right) }\overline{l}%
_{3L}\phi _{1}l_{3R}\frac{\chi ^{3}}{\Lambda ^{3}}+\varepsilon _{23}^{\left(
l\right) }\overline{l}_{2L}\phi _{1}l_{3R}\frac{\chi ^{3}}{\Lambda ^{3}}%
+\varepsilon _{22}^{\left( l\right) }\overline{l}_{2L}\phi _{1}l_{2R}\frac{%
\chi ^{5}}{\Lambda ^{5}}  \notag \\
&&+\varepsilon _{32}^{\left( l\right) }\overline{l}_{3L}\phi _{1}l_{2R}\frac{%
\chi ^{5}}{\Lambda ^{5}}+\varepsilon _{11}^{\left( l\right) }\overline{l}%
_{1L}\phi _{2}l_{1R}\frac{\chi ^{7}\zeta }{\Lambda ^{8}}  \notag \\
&&+\varepsilon _{11}^{\left( \nu \right) }\overline{l}_{1L}\widetilde{\phi }%
_{2}\nu _{1R}\frac{\chi ^{3}}{\Lambda ^{3}}+\varepsilon _{12}^{\left( \nu
\right) }\overline{l}_{1L}\widetilde{\phi }_{2}\nu _{2R}\frac{\chi ^{3}}{%
\Lambda ^{3}}  \notag \\
&&+\varepsilon _{21}^{\left( \nu \right) }\overline{l}_{2L}\widetilde{\phi }%
_{1}\nu _{1R}+\varepsilon _{22}^{\left( \nu \right) }\overline{l}_{2L}%
\widetilde{\phi }_{1}\nu _{2R}{+\varepsilon _{31}^{\left( \nu \right) }%
\overline{l}_{3L}\widetilde{\phi }_{1}\nu _{1R}}  \notag \\
&&{+\varepsilon _{32}^{\left( \nu \right) }\overline{l}_{3L}\widetilde{\phi }%
_{1}\nu _{2R}}+M_{1}\overline{\nu }_{1R}\nu _{1R}^{c}+M_{2}\overline{\nu }%
_{2R}\nu _{2R}^{c}  \notag \\
&&+M_{12}\overline{\nu }_{1R}\nu _{2R}^{c}+\varepsilon _{21}^{\left( \rho
\right) }\overline{l}_{2L}\phi _{1}\nu _{1R}\frac{\rho ^{+}}{\Lambda } 
\notag \\
&&+\varepsilon _{33}^{\left( l\right) }\overline{l}_{3L}\widetilde{\phi }%
_{1}l_{3R}\frac{\chi ^{3}\rho ^{+}}{\Lambda ^{4}}+\varepsilon _{23}^{\left(
l\right) }\overline{l}_{2L}\widetilde{\phi }_{1}l_{3R}\frac{\chi ^{3}\rho
^{+}}{\Lambda ^{4}}  \notag \\
&&+\varepsilon _{22}^{\left( \rho \right) }\overline{l}_{2L}\phi _{1}\nu
_{2R}\frac{\rho ^{+}}{\Lambda }{+\varepsilon _{31}^{\left( \rho \right) }%
\overline{l}_{3L}\phi _{1}\nu _{1R}}\frac{\rho ^{+}}{\Lambda }  \notag \\
&&{+\varepsilon _{32}^{\left( \rho \right) }\overline{l}_{3L}\phi _{1}\nu
_{2R}}\frac{\rho ^{+}}{\Lambda }+h.c  \label{lyl}
\end{eqnarray}

As the quark masses are related to the quark mixing parameters, we set the
vacuum expectation values (VEVs) of the SM singlet scalars with respect to
the Wolfenstein parameter $\lambda =0.225$ and the new physics scale $%
\Lambda $: 
\begin{equation}
v_{\xi }\sim v_{\zeta }\sim v_{\chi }=\lambda \Lambda .  \label{VEVsize}
\end{equation}

Regarding the Yukawa interactions of $\rho ^{+}
$ with quarks and leptons, we only consider operators up
to dimension eight and neglect higher dimensional contributions. From the quark Yukawa interactions, it follows that the
top partners will decay dominantly into either up or charm quarks
and the charged scalar singlet $\rho ^{+}$, whereas the dominant decay mode of $\rho ^{+}$ will
be into top and bottom quarks. Let us note that the
charged scalar singlet $\rho ^{+}$ cannot decay into charged leptons and
right handed neutrinos, since the right handed Majorana neutrinos are much
heavier than $\rho ^{+}$, thus not allowing that decay channel. In addition, the
charged scalar $\rho ^{+}$ can decay into charged leptons and light active
neutrinos but its corresponding decay rate is suppressed by $\lambda ^{6}%
\frac{v^{2}}{\Lambda ^{2}}$, as clearly seen from the lepton Yukawa
interactions. Consequently, the top partners can be searched at the LHC
through their decay channel $T_{m}\rightarrow \rho ^{+}u_{m}\rightarrow
tbu_{m}\rightarrow Wbbu_{m}\rightarrow l3j{\displaystyle{\not }}E_{T}$ ($m=1,2$). 
These top partners are produced in pairs at the LHC via a gluon fusion mechanism, where these exotic quarks are in the triangular loop followed by the propagator of the scalar $\chi $, followed again with a pair of the top partner and its
antiparticle (note that the scalar singlet $\chi $ has a renormalizable
coupling with these top partners). Thus observing an excess of events with
respect to the SM background in the opposite sign dileptons final state can
be a signal to confirm this model at the LHC.

From the Yukawa terms given above and considering that the VEV of $\xi $ is
aligned as $(1,0)$ in the $S_{3}$ direction \cite{Hernandez:2015dga}, we
find that the quark, charged lepton and light active neutrino mass matrices
are:

\begin{eqnarray}
M_{U} &=&\frac{v}{\sqrt{2}}\left( 
\begin{array}{ccc}
c_{1}\lambda ^{8} & 0 & a_{1}\lambda ^{3} \\ 
0 & b_{1}\lambda ^{4} & a_{2}\lambda ^{2} \\ 
0 & 0 & a_{3}%
\end{array}%
\right) ,\hspace{1cm}\hspace{1cm}  \notag \\
M_{D} &=&\frac{v}{\sqrt{2}}\left( 
\begin{array}{ccc}
e_{1}\lambda ^{7} & f_{1}\lambda ^{6} & 0 \\ 
e_{2}\lambda ^{6} & f_{2}\lambda ^{5} & 0 \\ 
0 & 0 & g_{1}\lambda ^{3}%
\end{array}%
\right) ,  \label{Mq}
\end{eqnarray}

\begin{eqnarray}
M_{l} &=&\frac{v}{\sqrt{2}}\left( 
\begin{array}{ccc}
x_{1}\lambda ^{8} & 0 & 0 \\ 
0 & y_{1}\lambda ^{5} & z_{1}\lambda ^{3} \\ 
0 & y_{2}\lambda ^{5} & z_{2}\lambda ^{3}%
\end{array}%
\right) ,  \label{Ml} \\
M_{\nu } &=&\left( 
\begin{array}{ccc}
W^{2} & \kappa WX & WY \\ 
\kappa WX & X^{2} & \kappa XY \\ 
WY & \kappa XY & Y^{2}%
\end{array}%
\right) ,\hspace{0.5cm}\kappa =\cos \varphi .  \notag
\end{eqnarray}

where $v=246$ GeV and $a_{k}$ ($k=1,2,3$), $b_{1}$, $c_{1}$, $g_{1}$, $f_{1}$%
, $f_{2}$, $e_{1}$ $e_{2}$, $x_{1}$, $y_{1}$, $y_{2}$, $z_{1}$, $z_{2}$ and $%
\kappa $ are $\mathcal{O}(1)$ parameters, whereas $X$, $Y$ and $W$\ are
parameters with dimension $\sqrt{m}$ where $m$ has mass dimension. The
Cabbibo mixing arises from the down-type quark sector whereas the up-type
quark sector contributes to the remaining mixing angles \cite%
{Hernandez:2014zsa}. Furthermore, light active neutrino masses arise via a
type I seesaw mechanism with two heavy right-handed Majorana neutrinos $\nu
_{1R}$ and $\nu _{2R}$. We have shown in \cite{Hernandez:2015dga} that the
fermion mass textures given above are consistent wih the current data on SM
fermion masses and mixings.

\textbf{The $750~\text{GeV}$ scalar resonance}. The recently reported excess
in the diphoton final state can be attributed to the $Z_{14}$ breaking
scalar $\chi $, which, taking into account the heavy exotic fermions, is
predominantly produced via gluon fusion through the triangular loop diagrams
with $T_{1}$ and $T_{2}$. The corresponding total cross section $\sigma $ is
a function of the gluon production rate $\Gamma (gg\rightarrow \chi )$ and
the consequent decay rate into photons $\Gamma (\gamma
\gamma \rightarrow \chi)$ 
\begin{align}
\Gamma (gg\rightarrow \chi )& =K^{gg}\frac{\alpha _{s}^{2}m_{\chi }^{3}}{%
32\pi ^{3}v_{\chi }^{2}}\bigg\vert F(x_{T})\bigg\vert^{2}, \\
\Gamma (\gamma \gamma \rightarrow \chi )& =\frac{\alpha ^{2}m_{\chi }^{3}}{%
64\pi ^{3}v_{\chi }^{2}}\bigg\vert N_{c}Q_{T}^{2}F(x_{T})\bigg\vert^{2},
\end{align}%
where $m_{\chi }\simeq 750\,$GeV denotes the resonance mass, $x_{T}=4m_{T}^{2}/m_{\chi }^{2}$, $m_T=y_Tv_{\chi}$, and $K^{gg}\sim 1.5$ accounts for higher order
QCD corrections. F(x) is a loop function given by 
\begin{equation*}
F(x)=2\,x\big(1+(1-x)f(x)\big),\,f(x)=\big(\arcsin \sqrt{1/x}\big)^{2}
\end{equation*}%
with $x_{T}>1\Leftrightarrow 4m_{T}>m_{\chi }$. Finally, we obtain 
\begin{equation*}
\sigma =\frac{\pi ^{2}}{8}\frac{\Gamma ( \gamma \gamma \rightarrow \chi)%
\frac{1}{s}\int_{m_{\chi }^{2}/s}^{1}\frac{dx}{x}f_{g}(x)f_{g}\left( \frac{%
m_{\chi }^{2}}{sx}\right) \Gamma (gg\rightarrow \chi )}{m_{\chi }\Gamma
_{\chi }}
\end{equation*}%
with $\sqrt{s}=13\,$TeV being the LHC center of mass energy, $\Gamma _{\chi
} $ the total decay width of $\chi $ and $f_{g}(x)$ the gluon distribution
function. To obtain a rough estimate of $\sigma $ we assume for simplicity
unified and natural Yukawa couplings of the exotic quarks $y_{T_{i}}\sim 1$,
which with $v_{\chi }\approx 1.2\,$TeV amounts to $\sigma \approx 8\,$fb.
This is well within the limits given by the ATLAS and CMS experiments \cite%
{Franceschini:2015kwy} 
\begin{equation*}
\sigma _{\text{ATLAS}}=10\pm 3\,\text{fb}\,,\quad \sigma _{\text{CMS}}=6\pm
3\,\text{fb}\,.
\end{equation*}%

\begin{figure}[t]
\subfigure{\includegraphics[width=0.45\textwidth]{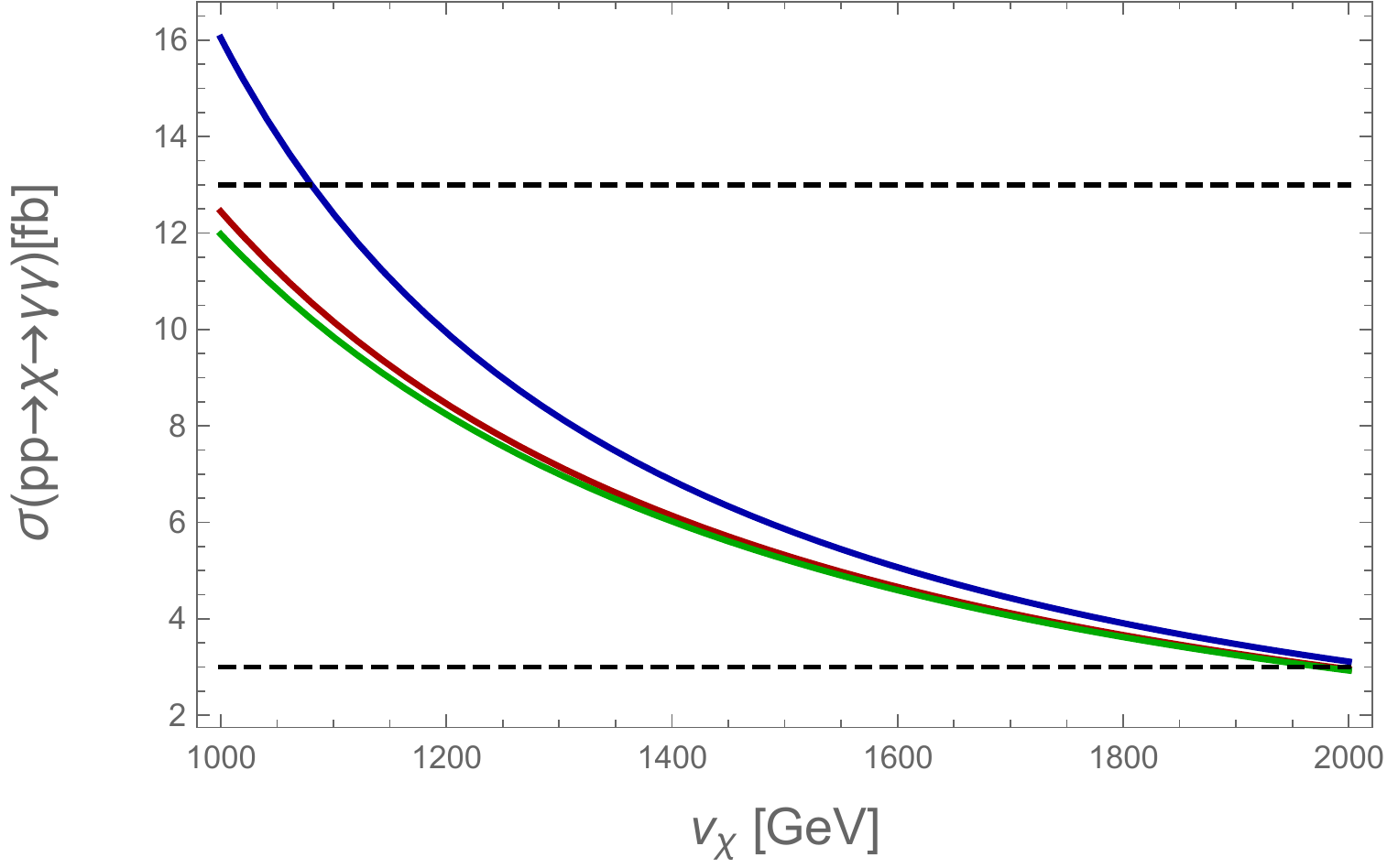}} 
\raggedright
\caption{Total cross section $\protect\sigma(pp \rightarrow \protect\chi %
\rightarrow \protect\gamma\protect\gamma)$ as a function of $v_{\protect\chi %
}$ for and different values of the exotic quark Yukawa couplings $y_{T_i} \sim 0.5$, $1$
and $1.5$ (in the curves from top to bottom, respectively), assuming $%
\protect\sqrt{s}=13\,TeV$ and $\protect\alpha _{s}(m_{\protect\chi %
}/2)\simeq 0.1$. The horizontal lines denote the experimentally allowed
limits of the diphoton signal given by ATLAS and CMS, which amount to $%
10\,\pm 3fb$ and $6\pm 3\,fb$, respectively. The limits require $v_{\protect%
\chi} \lesssim 2\,\text{TeV}$ if natural order one exotic quark Yukawa
couplings are assumed.}
\label{Figure1}
\end{figure}

The total cross section was computed using the MSTW2008
next-to-leading-order gluon distribution functions \cite{Martin:2009iq} as a
function of $v_{\chi}$ for different values of the exotic quark Yukawa
couplings. As shown in Fig. \ref{Figure1}, the cross section depends
crucially on the VEV $v_{\chi}$ as well as on the Yukawa couplings, which 
if sizable can also enhance $\sigma$ significantly in particular
for lower $v_{\chi}$ values.

If we further require that $\sigma $ be within the experimental limits given
by ATLAS and CMS, we predict $v_{\chi }$ to be smaller than $2\,$TeV, which
on the one hand sets the $Z_{14}$ breaking scale and on the other hand fixes
the expected particle masses of $\chi $ and the exotic quarks $T_{i}$ to be
in the same region.

\textbf{Conclusion}. The same flavon that is responsible for the shaping of
the fermion mass and mixing matrices can explain the recently reported $%
750\, $GeV excess in the diphoton channel. This is shown using a predictive
flavor model based on the $S_{3}\otimes Z_{3}\otimes Z_{3}^{\prime }\otimes
Z_{14}$ symmetry with the addition of heavy exotic fermions with electric
charge $\frac{5}{3}$ and a electrically charged scalar singlet. These heavy exotic quarks decay into a charged scalar singlet and the SM up type quarks, whereas the charged scalar singlet will mainly decay into SM up and down type quarks. Attributing the $Z_{14}$ breaking scalar to the
resonance allows one to fix the energy of the breaking scale and, hence,
enables immediate testing of the model at the current LHC run.


\textbf{Acknowledgments}. This project has received funding from the
European Union's Seventh Framework Programme for research, technological
development and demonstration under grant agreement no PIEF-GA-2012-327195
SIFT.

\end{document}